\def\nk{n_{\rm b}}

\def\Pb{P_{\rm b}}

\def\rfr#1{Equation~(\ref{#1})}
\def\rfrs#1#2{Equations~(\ref{#1})-(\ref{#2})}

\def\dert#1#2{\frac{{{\mathrm{d}}}{#1}}{{{\mathrm{d}}}{#2}}}

\def\virg#1{``#1"}

\def\eqi{\begin{equation}}
\def\eqf{\end{equation}}
\def\eqia{\begin{eqnarray}}
\def\eqfa{\end{eqnarray}}

\def\rp#1#2{{#1\over#2}}
\def\lb#1{\label{#1}}

\def\bds#1{\boldsymbol{#1}}

\def\ton#1{\left(#1\right)}
\def\qua#1{\left[#1\right]}
\def\grf#1{\left\{#1\right\}}

\documentclass[onecolumn]{aastex}
\usepackage{morefloats}
\usepackage[title]{appendix}
\usepackage{hyperref}
\usepackage{booktabs}
\usepackage[table,xcdraw]{xcolor}
\usepackage{multirow}
\usepackage{rotating,tabularx}
\usepackage{float}
\usepackage{enumerate}
\usepackage{rotating}
\usepackage[polutonikogreek,english]{babel}
\usepackage{amsmath,starfont,textgreek,w-greek,wasysym}
\usepackage{amsthm}
\usepackage{amscd,lineno}
\usepackage{amssymb,dsfont}
\usepackage{graphicx,epsfig}
\usepackage{txfonts}
\usepackage{xr-hyper}

\RequirePackage{color}

\newcommand{\emaila}{lorenzo.iorio@libero.it}

\allowdisplaybreaks[1]

\begin{document}

\title{What Would Happen If We Were About 1 pc Away from a Supermassive Black Hole?}
\shortauthors{L. Iorio}
\author{Lorenzo Iorio\altaffilmark{1} }
\affil{Ministero dell'Istruzione, dell'Universit\`{a} e della Ricerca
(M.I.U.R.),
\\ Viale Unit\`{a} di Italia 68, I-70125, Bari (BA),
Italy}
\email{\emaila}

\begin{abstract}
We consider a hypothetical planet with the same mass $m$, radius $R$, angular momentum $\boldsymbol{S}$, oblateness $J_2$, semimajor axis $a$, eccentricity $e$, inclination $I$, and obliquity $\varepsilon$ of the Earth orbiting a main-sequence star with the same mass $M_\star$ and radius $R_\star$ of the Sun at a distance $r_\bullet \simeq 1\,\mathrm{parsec}\,\ton{\mathrm{pc}}$ from a supermassive black hole in the center of the hosting galaxy with the same mass $M_\bullet$ of, say, $\mathrm{M87}^\ast$. We preliminarily investigate some dynamical consequences of its presence in the neighborhood of such a stellar system on the planet's possibility of sustaining complex life over time.  In particular, we obtain general analytic expressions for the long-term rates of change, doubly averaged over both the planetary and the galactocentric orbital periods $\Pb$ and $P_\bullet$, of $e,\,I,\,\varepsilon$, which are the main quantities directly linked to the stellar insolation. We find that, for certain orbital configurations, the planet's perihelion distance $q=a\ton{1-e}$  may  greatly shrink and even lead to, in some cases, an impact with the star. $I$ may also notably change, with variations even of the order of tens of degrees. On the other hand, $\varepsilon$ does not seem to be particularly affected, being shifted, at most, by $\simeq 0^\circ.02$ over $1$ Myr. Our results strongly depend on the eccentricity $e_\bullet$ of the galactocentric motion.
\end{abstract}

Unified Astronomy Thesaurus concepts:{
Black holes (162); Exoplanets (498); Celestial mechanics (211);
Astrobiology (74); Gravitation (661);
}

\section{Introduction}\lb{intro}
Let us consider a restricted two-body system consisting of a terrestrial planet orbiting a main-sequence star which, in turn, revolves around a supermassive black hole (SMBH) along an orbit whose perinigricon\footnote{This is one of the possible names attributable  to the pericentre when the primary is a black hole \citep{2002Natur.419..694S}. It comes from the Latin word \virg{\textit{niger}}, meaning \virg{black}.} distance amounts to hundreds or thousands of Schwarzschild radii. For the sake of definiteness, in the following we will assume the same physical and orbital parameters of the Sun and the Earth, while we will adopt the mass $M_\bullet =6.5\times 10^9\,M_\odot$ \citep{2019ApJ...875L...6E} of M87$^\ast$ located in the center of the supergiant elliptical galaxy M87.
Its shadow was recently imaged for the first time \citep{2019ApJ...875L...1E}. Notably, even more massive SMBHs exist like, e.g., TON 618 $\ton{M_\bullet=6.6\times 10^{10}\,M_\odot}$ \citep{2004ApJ...614..547S}, and HOLM 15A $\ton{M_\bullet=4.0\times 10^{10}\,M_\odot}$ \citep{2019arXiv190710608M}.

A scenario like the aforementioned one should not be deemed either entirely fictional or unrealistic. It has been recently demonstrated that the formation of many (rogue) Earth-sized planets in the circumnuclear disks around SMBHs in low luminosity active galactic nuclei (AGNs) is potentially possible at several parsecs (pc) from them \citep{2019arXiv190906748W}.
Several pros and cons of the habitability of such kind of worlds  have been studied by \citet{2017AmJPh..85...14O}, \citet{2018MNRAS.479..171F}, \citet{Lingam_2019}, and \citet{2019arXiv191000940S}.
In particular, \citet{Lingam_2019} analyzed the strong radiation environment in AGNs hosting SMBHs and found that, perhaps, they may be not so extremely hazardous for life. Indeed, the zone where the negative effects due to strong ultraviolet (UV) irradiation become dominant seems to be smaller than the corresponding zones for powering prebiotic chemistry and photosynthesis, which may extend up to $\simeq 40-340\,\mathrm{pc}$. On the other hand, close-in planets at $\lesssim 1\,\mathrm{pc}$ might become uninhabitable because of complex interactions with the dusty torus as well as strong outflows and winds at relativistic speeds from the accretion disk. For a study quantifying the extent to which the activity of $\mathrm{Sgr\,A}^\ast$ may have affected the habitability of Earth-like planets in our Galaxy, see \citet{2017NatSR...716626B}, which mainly focuses on the effects of electromagnetic irradiation produced during the peak of the active phase on the atmospheres of terrestrial planets.
Stellar formation in the Galactic center (GC) was investigated by, e.g., \citet{1993ApJ...408..496M}, \citet{kauffmann_2016}, and \citet{Kruij2019}. It should be recalled that 8 out of the roughly 40 stars constituting the S-star cluster orbiting $\mathrm{Sgr\,A}^\ast$ in the central arcsecond of the GC are old,  main-sequence stars of spectral classes G,\,K,\,and M whose masses are in the range $0.5-2\,M_\odot$ \citep{2019arXiv190806298D,2019ApJ...872L..15H}.
The evolution of a fictitious planetary system like our solar system around an S-type star orbiting the SMBH in $\mathrm{Sgr\,A}^\ast$ was recently studied by \citet{2019arXiv190806298D}, who recalled that the existence of planetary systems in the innermost parsec of $\mathrm{Sgr\,A}^\ast$ is still debated \citep{2016ApJ...831...61T}, perhaps because of migration instead of in situ formation.
Finally, we also mention in passing the intriguing possibility that advanced civilizations may develop inside SMBHs themselves \citep{2011CQGra..28w5015D,2012GrCo...18...65D}.

We wish to preliminarily investigate some dynamical features of motion of a Sun--planet two-body system with an  SMBH in its relatively close neighborhood. In fact, such features have a general validity, not being necessarily restricted to the considered scenario. Nonetheless, in the present case, they  may have relevant consequences on the potential for the evolution of intelligent life forms on the planet, and even on the very same existence of the planet itself for certain orbital configurations. In particular, we will look at the long-term evolution of the planet's orbital eccentricity $e$ and inclination $I$, and of the axis of  spin angular momentum $\hat{S}$. Indeed, it is also known that the inclination $I$ of the Earth's orbit affects incoming solar irradiance \citep{Vieira2012} and, thus, its climate and habitability, not to mention the role played by eccentricity and obliquity $\varepsilon$, i.e. the inclination of the Earth's equator to the ecliptic, in affecting the Earth's insolation over time \citep{1993A&A...270..522L,1993Natur.361..615L,1997Icar..129..254W,2004A&A...428..261L,2002IJAsB...1...61W,2010ApJ...721.1295D,2010ApJ...721.1308S,2015P&SS..105...43L,2017ApJ...844..147K}. Given the scenario considered, in addition to the high energy radiation emission coming from the star,  the emission from the matter surrounding the SMBH itself during the various phases of the orbital evolution should be taken into account in the assessment of the overall planet's habitability. A straightforward, although simplistic to a certain extent,  way to quantify such an aspect would consist of a comparison of the flux arriving on the planet from the star with the flux from the AGN or accretion disk; this is outside the scope of the present paper.
In order to get easily some broad insights from analytical calculations, we will adopt a simplified model which, however, under certain circumstances, may be extended also to an $N$-body solar system like ours. To be more specific, we will assume that possible variations of the other orbital elements of the planet and of its spin axis induced by the gravitational interaction with other possible members of its planetary system like, e.g., planets and Moon-like satellite(s) and its parent star (stellar oblateness, post-Newtonian effects) are  characterized by timescales much longer than the orbital period around the SMBH. In the case of Earth, the precession of the equinoxes occurs in $0.086\,\mathrm{Myr}$, while its orbital elements change in $\simeq 0.1-3\,\mathrm{Myr}$ \citep[Tab.\,A.3]{2000ssd..book.....M}. It should be kept in mind that we are not embarking on a detailed paleoclimatological or archeoastronomical investigation of a specific existing habitat like Earth; instead, we are just interested in outlining a general picture that captures some salient aspects and that, hopefully, encourages further, more detailed studies. Finally, let us note that, strictly speaking, our analysis may not necessarily be limited just to an SMBH, but it may be valid, to a certain extent, for any (ordinary or possibly exotic) matter concentration of identical mass enclosed in a spatial volume with a size larger than the corresponding Schwarzschild radius. As far as, say,  global clusters are concerned, it seems unlikely that they can support the formation of terrestrial, Earth-like planets for various reasons \citep{2001Icar..152..185G,2018ApJ...864..115K}. Moreover, the typical sizes of the  globular clusters range from about $20-100$ pc or more, with masses as little as
$\simeq 10^4-10^5\,M_\odot$ \citep{bookglobular}\footnote{See also https://www.astro.keele.ac.uk/workx/globulars/globulars.html.} . Even if they had the same mass of the SMBH considered here,  the putative Sun--Earth system under consideration should be much more distant in order to gravitationally consider them as equivalent to  distant point masses, thus weakening the dynamical effects considered here.

The paper is organized as follows. In Section\,\ref{keple}, we analytically work out, in a perturbative way, the doubly averaged long-term rates of change of the semimajor axis $a$, the eccentricity $e$, the inclination $I$ to the reference $\grf{x,\,y}$ plane, the longitude of the ascending node $\Omega$, and the longitude of periastron $\varpi$ of the Earth-like planet. Furthermore, we discuss the evolution of the periastron distance $q=a\ton{1-e}$ and $I$ over a $\simeq \mathrm{Myr}$ timescale or so. Section\,\ref{spinprec} deals with the long-term rates of change of the components of the planet's spin axis $\hat{S}$, with particular emphasis on the obliquity $\varepsilon$. Section\,\ref{fine} summarizes our findings and offers our conclusions.
\section{Orbital Perturbations due To a Distant SMBH}\lb{keple}
Here, we will analytically calculate the long-term orbital perturbations induced by the distant SMBH on the motion of the Earth-like planet around its Sun-type parent star by doubly averaging them over both the year-long planetary orbital period $\Pb$ and the much longer period $P_\bullet$ of the revolution of the Sun--planet system around the SMBH itself. In performing the second average, we will assume that any possible long-term variations of the planetary orbital parameters due to, e.g., the putative $N$-body interactions with any other planets orbiting the same star and the classical and relativistic post-Keplerian (pK) components (stellar oblateness $J_2^\star$, Schwarzschild, Lense-Thirring effect) of the gravitational field of the latter occur much more slowly than the galactic revolution itself.
As an example, let us note that the periods of the overall changes experienced by the inclination $I_\oplus$, the longitude of the ascending node $\Omega_\oplus$, and the longitude of perihelion $\varpi_\oplus$ of the real Earth, inferable from the their secular rates of change as released by the HORIZONS Web-Interface, maintained by the NASA Jet Propulsion Laboratory (JPL), are as long as $\simeq 0.1-3\,\mathrm{Myr}$; see also \citet[Tab.\,A.3]{2000ssd..book.....M}. On the other hand, a galactocentric semimajor axis as large as, say, $a_\bullet \simeq 1\,\mathrm{pc}$, corresponding to a perinigricon distance of some $1500$ Schwarzschild radii\footnote{The Schwarzschild radius of $\mathrm{M}87^\ast$ amounts to $R_\mathrm{S} = 2GM_\bullet/c^2\simeq 128\,\mathrm{au}$, where $G$ is the Newtonian constant of gravitation and $c$ is the speed of light in vacuum. } $\ton{R_\mathrm{S}}$ for moderate eccentricities $e_\bullet\lesssim 0.1$ and of $\simeq 500-150\,R_\mathrm{s}$ for $e_\bullet\simeq 0.7-0.9$, gives a revolution period $P_\bullet \simeq 0.001\,\mathrm{Myr}=1\,\mathrm{kyr}=10\,\mathrm{cty}$. Incidentally, such a distance from the SMBH assures that the planet would not be tidally locked, as per Figure\,2 of \citet{2019arXiv191000940S}, thus avoiding the related consequences on its habitability. Moreover, if we assume  values of the order of $\sigma\simeq 1-1.5\times 10^{-3}\,c$ \citep{2014ApJ...785..143M} for the stellar velocity dispersion $\sigma$, as for M$87^\ast$, the orbit of our fictitious star--planet system falls well within the SMBH's sphere of influence defined by $r_\mathrm{H}=GM_\bullet/\sigma^2\simeq 100-300\,\mathrm{pc}$  \citep{1972ApJ...178..371P}. Thus, we can neglect the effects of the whole galactic potential on the stellar trajectory which, otherwise, may turn out to be remarkably non-Keplerian \citep{1987gady.book.....B,Contopoulos:2011}.

Our calculation, to the quadrupole order of the tidal perturbing potential \citep{1991AJ....101.2274H} of a distant, point-like perturber, is not restricted to any particular orbital geometries of both the planetary and the galactic motions, i.e., our resulting formulas hold for any values of the eccentricities $e,\,e_\bullet$ and the inclinations $I,\,I_\bullet$ of the astrocentric and galactocentric orbits, respectively. As such, our results have a general validity, and hold in any coordinate system.

Inserting the doubly averaged\footnote{For the reasons explained above, we can assume Keplerian ellipses as reference trajectories in both averages.} quadrupole tidal potential due to a point-like, massive distant perturber \citep{1991AJ....101.2274H} in the Lagrange planetary equations \citep{2000ssd..book.....M,Bertotti03,2011rcms.book.....K} allows us to straightforwardly obtain
\begin{align}
\dert a t \lb{adot}& = 0, \\ \nonumber \\
\dert e t \nonumber \lb{edot}& = \rp{15\,GM_\bullet\,e\,\sqrt{1-e^2}}{32\,a_\bullet^3\,\ton{1-e^2_\bullet}^{3/2}\,\nk}\grf{\qua{\ton{1 + 3\,\cos\,2I_\bullet}\,\sin^2\,I + \ton{3 + \cos\,2I}\,\cos\,2\Delta\Omega\,\sin^2\,I_\bullet -\right.\right.\\ \nonumber \\
\nonumber &-\left.\left. 2\,\cos\Delta\Omega\,\sin\,2I\,\sin\,2I_\bullet}\,\sin\,2\omega + 8\,\cos\,2\omega\,\sin\,I_\bullet\,\qua{-\cos\,I_\bullet\,\sin\,I +\right.\right.\\ \nonumber \\
&+\left.\left. \cos\,I\,\cos\Delta\Omega\,\sin\,I_\bullet} \sin\Delta\Omega },\\ \nonumber \\
\dert I t \nonumber \lb{Idot}& =\rp{3\,GM_\bullet}{8\,a_\bullet^3\,\ton{1-e^2_\bullet}^{3/2}\,\sqrt{1-e^2}\,\nk}\ton{\cos I\,\cos I_\bullet + \cos\Delta\Omega\,\sin I\,\sin I_\bullet}\times\\ \nonumber \\
\nonumber &\times \qua{5\,e^2\,\ton{-\cos I_\bullet\,\sin I +\cos I\,\cos\Delta\Omega\,\sin I_\bullet}\,\sin 2\omega + \right.\\ \nonumber \\
&+\left. \ton{2 + 3\,e^2 + 5\,e^2\,\cos 2\omega}\,\sin I_\bullet\,\sin\Delta\Omega}, \\ \nonumber \\
\dert \Omega t \nonumber \lb{Odot}& = -\rp{3\,GM_\bullet}{8\,a_\bullet^3\,\ton{1-e^2_\bullet}^{3/2}\,\sqrt{1-e^2}\,\nk}\ton{\cos I_\bullet\,\cot I + \cos\Delta\Omega\,\sin I_\bullet}\times\\ \nonumber \\
\nonumber &\times \qua{-\ton{-2 - 3\,e^2 + 5\,e^2\,\cos 2\omega}\ton{\cos I_\bullet\,\sin I -
\cos I\,\cos\Delta\Omega\,\sin I_\bullet} - \right.\\ \nonumber \\
&-\left. 5\,e^2 \sin I_\bullet\,\sin 2\omega\,\sin\Delta\Omega} , \\ \nonumber \\
\dert \omega t  \lb{odot}\nonumber & = -\rp{GM_\bullet}{8\,a_\bullet^3\,\ton{1-e^2_\bullet}^{3/2}\,\sqrt{1-e^2}\,\nk}\grf{3\,\cot I\,\ton{\,\cos I\,\cos I_\bullet +\cos \Delta\Omega\,\sin I\,\sin I_\bullet}\times\right.\\ \nonumber \\
\nonumber &\left. \times \qua{\ton{-2 - 3 e^2 + 5 e^2\,\cos 2\omega}\,\ton{\cos I_\bullet\,\sin I - \cos I\,\cos \Delta\Omega\,\sin I_\bullet} + 5\,e^2\,\sin I_\bullet\,\sin 2\omega\,\sin \Delta\Omega} + \right.\\ \nonumber \\
\nonumber &\left. +\rp{3}{8}\,\ton{-1 + e^2}\,\grf{1 +\cos 2I_\bullet\,\ton{3 + 30\,\cos 2\omega\,\sin^2 I} + 6\,\cos 2\Delta\Omega\,\sin^2 I_\bullet + \right.\right.\\ \nonumber \\
\nonumber &\left.\left. + \cos 2I\,\qua{3 + 9\,\cos 2I_\bullet + 2\,\ton{-3 + 5\,\cos 2\omega}\,\cos 2\Delta\Omega\,\sin^2 I_\bullet} +\right.\right.\\ \nonumber \\
\nonumber & + \left.\left. 12\,\cos \Delta\Omega\,\sin 2I\,\sin 2I_\bullet + 10\,\qua{\cos 2\omega\,\ton{\sin^2 I + 3\,\cos 2\Delta\Omega\,\sin^2 I_\bullet -\right.\right.\right.\right.\\ \nonumber \\
&\left.\left.\left.\left. - 2\,\cos \Delta\Omega\,\sin 2I\,\sin 2I_\bullet} + 4\,\sin 2\omega\,\ton{-\,\cos I\,\sin I_\bullet^2\,\sin 2\Delta\Omega + \sin I\,\sin 2I_\bullet\,\sin \Delta\Omega}}} }.
\end{align}
In \rfrs{adot}{odot}, $\nk=\sqrt{GM_\star/a^3}=2\uppi/\Pb$ is the Keplerian mean motion of the planet's orbit around its star of mass $M_\star=1\,M_\odot$, while $\Delta\Omega\doteq \Omega-\Omega_\bullet$ is the difference of the nodes of the planetary and SMBH orbital planes, and $\omega$ is the planet's argument of periastron.

\begin{figure}[H]
\centering
\centerline{
\vbox{
\begin{tabular}{c}
\epsfysize= 7.0 cm\epsfbox{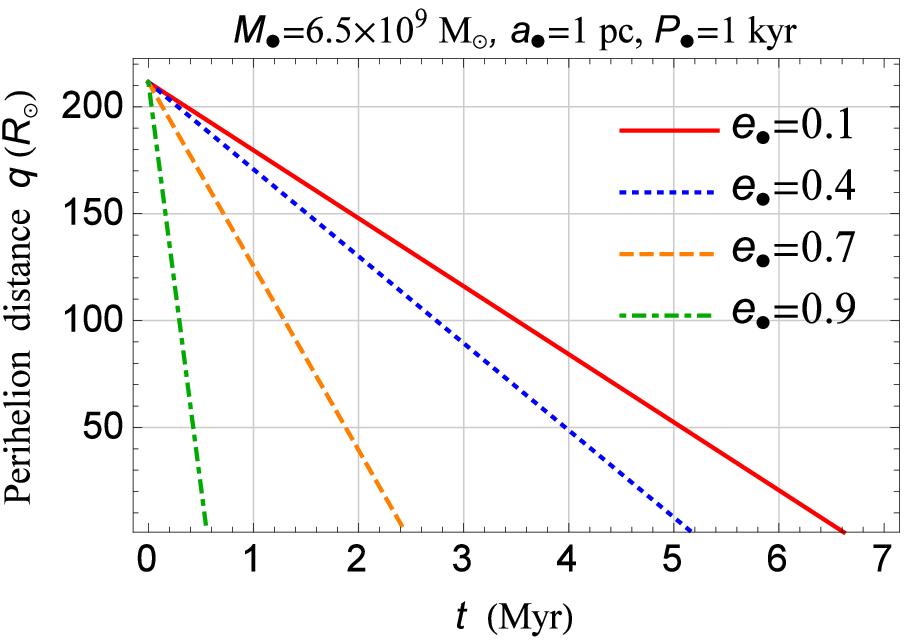}\\
\epsfysize= 7.0 cm\epsfbox{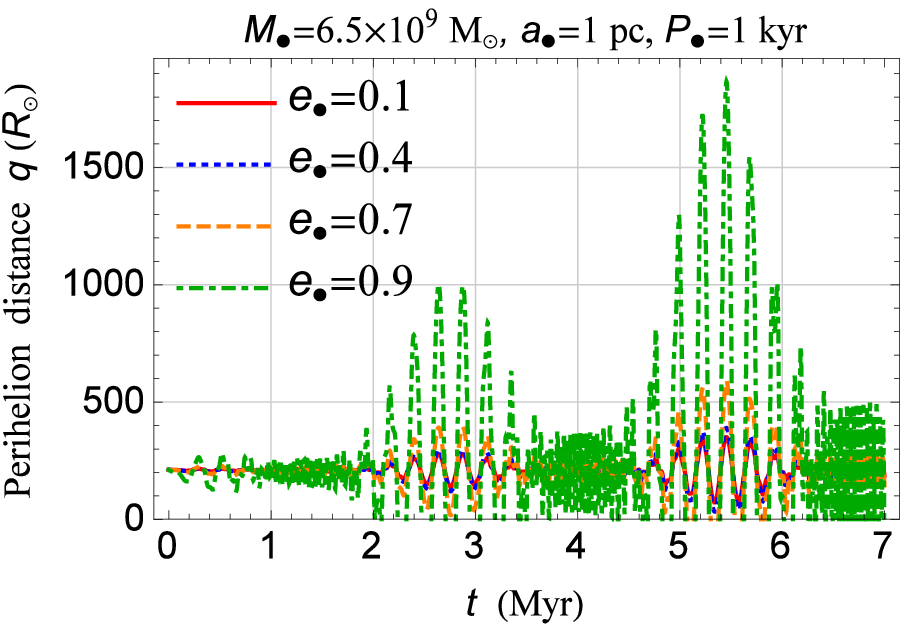}\\
\end{tabular}
}
}
\caption{
Upper panel: plot of the perihelion distance $q\ton{t}=a_0\ton{1-e_0-\Delta e\ton{t}}$, in units of solar radii $R_\odot$, as a function of time $t$, in millions of years, over a time span $\Delta t=7\,\mathrm{Myr}$ for different values of the eccentricity $e_\bullet$ of the orbital motion of the star--planet system around an SMBH with $M_\bullet = 6.5\times 10^9\,M_\odot$ occurring in $P_\bullet = 1\,\mathrm{kyr}\ \ton{a_\bullet = 1\,\mathrm{pc}}$. In order to obtain $\Delta e\ton{t}$, we straightforwardly integrated \rfr{edot} by keeping the astrocentric and galactocentric orbital parameters entering it fixed  to $a_0=1\,\mathrm{au},\,e_0=0.0167,\,I_0=0,\,\omega_0=151^\circ,\,I_\bullet=90^\circ$, and $\Delta\Omega_0=254^\circ$. The assumed values of the Euler-type angular variables maximize the right-hand side of \rfr{edot} which turns out to be positive. The intersection of the curves with the horizontal axis, marked by $q=1\,\mathrm{R}_\odot$, corresponds to the impact of the planet with its parent star. Lower panel: same as in the upper panel, apart from the fact that $\Delta e\ton{t}$ was integrated by assuming the same $N$-body rates of change for the planet's orbital elements as of the Earth. They were retrieved from the HORIZONS Web-Interface maintained by JPL, NASA; see also \citet[Tab.\,A.3]{2000ssd..book.....M}.
  }\label{figura1}
\end{figure}

In analogy with the Sun--Earth case, let us assume that the planetary orbital motion occurs just in the reference $\grf{x,\,y}$ plane adopted, i.e., we put $I=0$, and  focus on \rfr{edot}. It turns out that its trigonometric part enclosed in the curly brackets, which is characterized by the orientations of the orbital planes of the planet and of the SMBH determining, among other things, the sign of the rate of change of the eccentricity, assumes its maximum positive value, corresponding to 4, for $\Delta\Omega= 254^\circ,\,I_\bullet = 90^\circ,\,\omega=151^\circ$. In the following, we will adopt such values in order to maximize the effects we are interested in.
Figure\,\ref{figura1} deals with a system revolving around the SMBH within a time span of $P_\bullet = 1\,\mathrm{kyr}$  by depicting the temporal evolutions of the perihelion distance $q=a\ton{1-e}$ to its parent star over $\Delta t = 7\,\mathrm{Myr}$ for different values of the eccentricity $e_\bullet$ of the galactocentric orbit.
In the upper panel,  it was assumed that the no other pK variations of the planetary orbital elements occur or, equivalently, that their timescales are much longer than $P_\bullet$ and $\Delta t$. It can be noticed that, for the adopted orbital geometry, an Earth-type orbit starting at the same distance of our planet from the Sun soon becomes unable to sustain life because of the steady increase of the insolation due to the decrease of $q$, even ending on the star after few million years. The larger the eccentricity of the galactocentric orbit, the sooner the planet impacts its parent star.
In the lower panel, any $N$-body secular rates of change of the planet's orbital elements were taken into account by assuming the same values of the Earth for them. The picture is, now, quite different, showing wild harmonic variations, especially pronounced for large galactocentric eccentricities, which may even lead the planet to impact its star after a few million years. Anyway, the excursions turn out to be so huge that it seems unlikely that the long-term habitability on such a world would be preserved. Suffice to say that an inspection of Figure\,4 of \citet{Vieira2012} shows that the the maximum annually integrated change of the Earth's total solar irradiance (TSI) related to the orbital eccentricity variation of just $\Delta e\simeq 0.045$ over the last $0.6\,\mathrm{Myr}$ is as little as $\simeq 1.5\,\mathrm{W\,m}^{-2}$ \citep{2004A&A...428..261L}, i.e. roughly $\simeq 0.1\%$ of the TSI's baseline value of approximately $1.361\times 10^3\,\mathrm{W\,m}^{-2}$ during minima in solar activity \citep{2011GeoRL..38.1706K}.

\begin{figure}[H]
\centering
\centerline{
\vbox{
\begin{tabular}{c}
\epsfysize= 7.0 cm\epsfbox{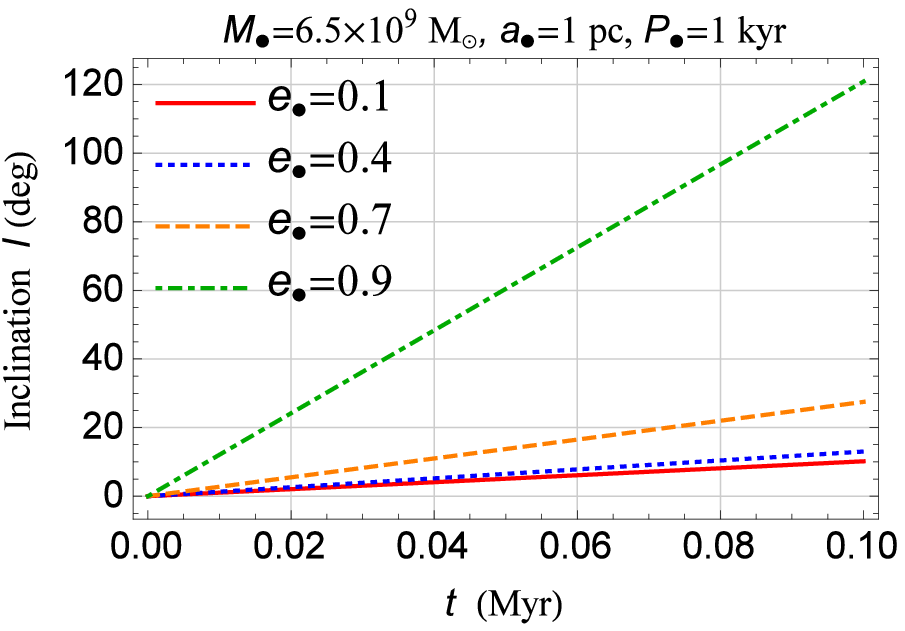}\\
\epsfysize= 7.0 cm\epsfbox{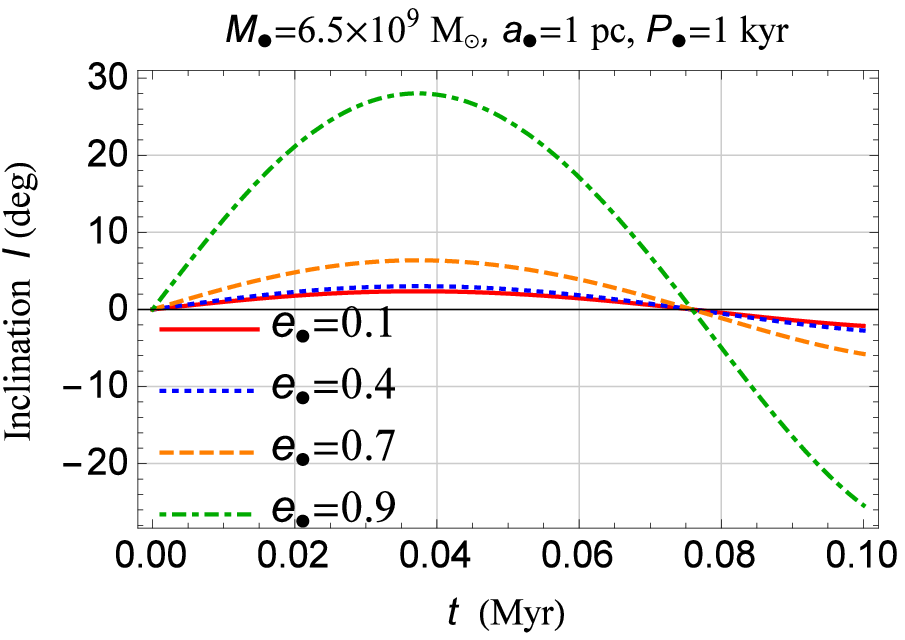}\\
\end{tabular}
}
}
\caption{
Upper panel: plot of the inclination $I$, in degrees, as a function of time $t$, in millions of years, over a time span $\Delta t=0.1\,\mathrm{Myr}$ for different values of the eccentricity $e_\bullet$ of the orbital motion of the star--planet system about a SMBH with $M_\bullet = 6.5\times 10^9\,M_\odot$ occurring in $P_\bullet = 1\,\mathrm{kyr}\ \ton{a_\bullet = 1\,\mathrm{pc}}$. In order to obtain $\Delta I\ton{t}$, we straightforwardly integrated \rfr{Idot} by keeping the astrocentric and galactocentric orbital parameters entering it fixed  to $a_0=1\,\mathrm{au},\,e_0=0.0167,\,I_0=0,\,\omega_0=180^\circ,\,I_\bullet=135^\circ$, and $\Delta\Omega_0=270^\circ$. The assumed values of the Euler-type angular variables maximize the right-hand side of \rfr{Idot} which turns out to be positive. Lower panel: same as in the upper panel, apart from the fact that $\Delta I\ton{t}$ was integrated by assuming the same $N$-body rates of change for the planet's orbital elements as of the Earth. They were retrieved from the HORIZONS Web-Interface maintained by JPL, NASA; see also \citet[Tab.\,A.3]{2000ssd..book.....M}.
  }\label{figura2}
\end{figure}

Figure\,\ref{figura2} displays the temporal evolutions of the planet's orbital inclination $I$  over $\Delta t = 0.1\,\mathrm{Myr}$ for different values of the eccentricity $e_\bullet$ of the galactocentric orbit. The upper panel is based on the straightforward integration of \rfr{Idot} by keeping the planet's perihelion fixed to the value $\omega=180^\circ$ which, along with $I_\bullet=135^\circ,\,\Delta\Omega=270^\circ$, yields the maximum of the trigonometric factor in the right-hand side of \rfr{Idot}. The resulting change of the inclination can be quite remarkable, especially for very eccentric motion around the SMBH.
The lower panel of Figure\,\ref{figura2} takes into account putative $N$-body interactions with any other planets of the stellar system assuming them equal to those of the Earth, as in Figure\,\ref{figura1}. Also in this case, the amplitudes of the resulting harmonic shifts can reach tens of degrees for $e_\bullet=0.7-0.9$, at least over the time span adopted.
For the sake of a comparison, according to Figure\,4 of \citet{Vieira2012},  for rather modest variations of $I_\oplus$, the Earth's TSI changes by up to $0.14\,\mathrm{W\,m}^{-2}$. However, variations of the Earth's orbital inclination during the last $0.6\,\mathrm{Myr}$ amounted to $\Delta I_\oplus\lesssim 2.5^\circ$ \citep{2003ApJ...592..620V}. In this scenario, the maximum terrestrial TSI modulation due to orbital inclination is $\simeq 3\times 10^{-3}\,\mathrm{W\,m}^{-2}$. Thus, it is arguable that the distant SMBH, if located at certain positions in the sky, may have a serious impact on the habitability of the considered Earth-like planet also through the perturbations on its inclination.

Finally, let us mention that, although not directly related to the stellar insolation, also the precessions of the node and the periastron may potentially have a somewhat indirect impact on the long-term habitability of the planet. Indeed, investigations on the so-called galactic habitability zone \citep{2008SSRv..135..313P} revealed that the central regions may be potentially harmful for life because of an enhanced supernovae (SNe) explosions, which are expected to be more numerous in the inner regions of galaxies. Thus, we believe that a rapidly varying position of the orbit in its orbital plane and of the orbital plane itself, caused by a fast precessing periastron and node, respectively, might somewhat augment the potential exposure of the planet to sterilizing SN events \citep{2017NatSR...7.5419S} over time.
\section{The Precession of the Spin Angular Momentum of the Planet}\lb{spinprec}
Here, we look at  possible long-term variations of the obliquity $\varepsilon$ of the planet's spin axis $\hat{S}$ with respect to the plane of its orbital motion around its parent star due to the gravitational pull of the SMBH. In order to describe the external torque acting on the planetary angular momentum, we will assume a simplified model, adequate for our illustrative scope. Indeed, it should be recalled once again that we are not carrying out any detailed paleoclimatological study of the real Earth,  instead our aim is to outline a general picture capable of capturing some key features of the investigated scenario.

By assuming that the Earth-like planet under consideration has mass $m$, equatorial radius $R$, spin angular momentum $\boldsymbol{S}$,  and dimensionless quadrupole mass moment  $J_2$,
the rate of change of $\boldsymbol{S}$ induced by the SMBH located in the direction $\bds{\hat{r}}_\bullet$ at distance $r_\bullet$ can be expressed as \citep{2014grav.book.....P}
\eqi
\dert{\bds{S}}t = -\rp{3\,G\,m\,J_2\,R^2\,M_\bullet}{r_\bullet^3}\,\ton{\bds{\hat{S}}\bds\cdot\bds{\hat{r}}_\bullet}\,\ton{\bds{\hat{S}}\bds\times\bds{\hat{r}}_\bullet}.
\eqf
Thus, the long-term precessions of the components ${\hat{S}}_x,\,{\hat{S}}_y,\,{\hat{S}}_z$ of the planet's spin axis $\bds{\hat{S}}$, averaged over one full galactocentric orbital period $P_\bullet$, turn out to be
\begin{align}
\dert{\hat{S}_x}t \lb{dsxdt} \nonumber & = \rp{3\,G\,m\,J_2\,R^2\,M_\bullet}{2\,S\,a^3_\bullet\,\ton{1-e^2_\bullet}^{3/2}}\ton{\hat{S}_y\,\cos I_\bullet +\hat{S}_z\,\sin I_\bullet\,\cos\Omega_\bullet}\qua{\hat{S}_z\,\cos I_\bullet +\right.\\ \nonumber \\
&\left. + \sin I_\bullet\,\ton{\hat{S}_x\,\sin\Omega_\bullet -\hat{S}_y\,\cos\Omega_\bullet } }, \\ \nonumber \\
\dert{\hat{S}_y}t \lb{dsydt} \nonumber & = -\rp{3\,G\,m\,J_2\,R^2\,M_\bullet}{2\,S\,a^3_\bullet\,\ton{1-e^2_\bullet}^{3/2}}\ton{\hat{S}_x\,\cos I_\bullet -\hat{S}_z\,\sin I_\bullet\,\sin\Omega_\bullet}\qua{\hat{S}_z\,\cos I_\bullet +\right.\\ \nonumber \\
&\left. + \sin I_\bullet\,\ton{\hat{S}_x\,\sin\Omega_\bullet -\hat{S}_y\,\cos\Omega_\bullet } }, \\ \nonumber \\
\dert{\hat{S}_z}t \lb{dszdt} \nonumber & = -\rp{3\,G\,m\,J_2\,R^2\,M_\bullet\,\sin I_\bullet}{2\,S\,a^3_\bullet\,\ton{1-e^2_\bullet}^{3/2}}\ton{\hat{S}_x\,\cos\Omega_\bullet +\hat{S}_y\,\sin\Omega_\bullet}\qua{\hat{S}_z\,\cos I_\bullet + \right.\\ \nonumber \\
&\left. + \sin I_\bullet\,\ton{\hat{S}_x\,\sin\Omega_\bullet -\hat{S}_y\,\cos\Omega_\bullet } }.
\end{align}
It should be noted that \rfrs{dsxdt}{dszdt}, which are valid for an arbitrary  SMBH's orbital configuration in any coordinate system, hold if ${\hat{S}}_x,\,{\hat{S}}_y,\,{\hat{S}}_z$, which were kept fixed in the integration over $P_\bullet$, can be considered as approximately constant during an orbital period of the galactocentric motion. Such a condition would be fulfilled in the case of our Earth since the lunisolar precession of the equinoxes occurs in $\simeq 26\,\mathrm{kyr}$, while $P_\bullet=1\,\mathrm{kyr}$.
By posing
\begin{align}
\hat{S}_x & = \sin\varepsilon\cos\beta, \\ \nonumber \\
\hat{S}_y & = \sin\varepsilon\sin\beta, \\ \nonumber \\
\hat{S}_z & = \cos\varepsilon,
\end{align}
and defining
\eqi
\Gamma\doteq\beta-\Omega_\bullet,
\eqf
\rfr{dszdt} yields for the spin's obliquity $\varepsilon$
\eqi
\dert\varepsilon t =\rp{3\,G\,m\,J_2\,R^2\,M_\bullet\,\sin I_\bullet\,\cos\Gamma}{2\,S\,a^3_\bullet\,\ton{1-e^2_\bullet}^{3/2}}
\ton{\cos I_\bullet\,\cos\varepsilon- \sin I_\bullet\,\sin\varepsilon\,\sin\Gamma}.\lb{obli}
\eqf
We remark that \rfrs{dsxdt}{dszdt}, and, in particular, \rfr{obli}, in view of their generality, can also be used in, e.g., the search for the hypothesized Telisto/Planet Nine \citep{2019PhR...805....1B}, putatively residing in the outskirts of our solar system, by looking, among other things, at its impact on the spin axis of the Sun \citep{2016AJ....152..126B}, and of the other known planets as well.

\begin{figure}[H]
\centering
\centerline{
\vbox{
\begin{tabular}{c}
\epsfysize= 7.0 cm\epsfbox{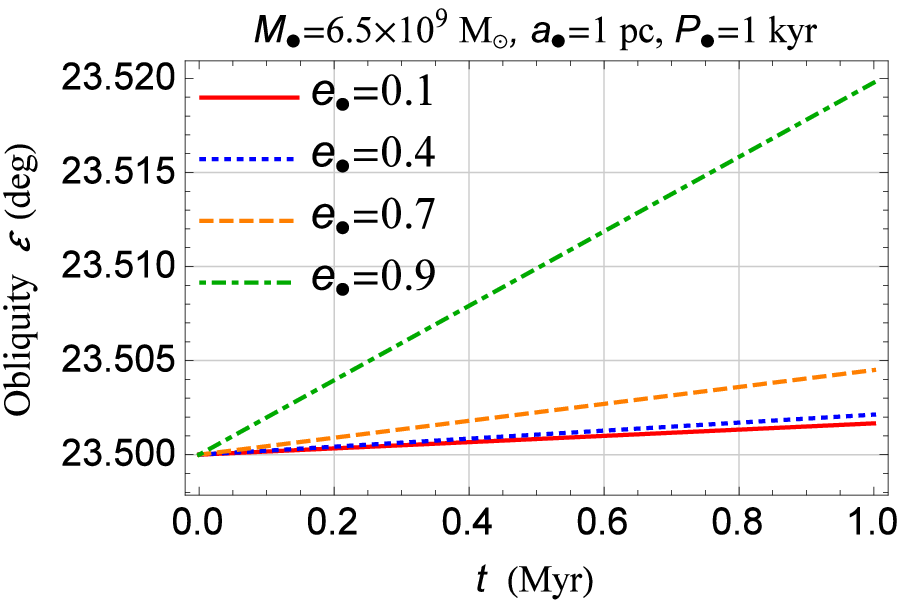}\\
\epsfysize= 7.0 cm\epsfbox{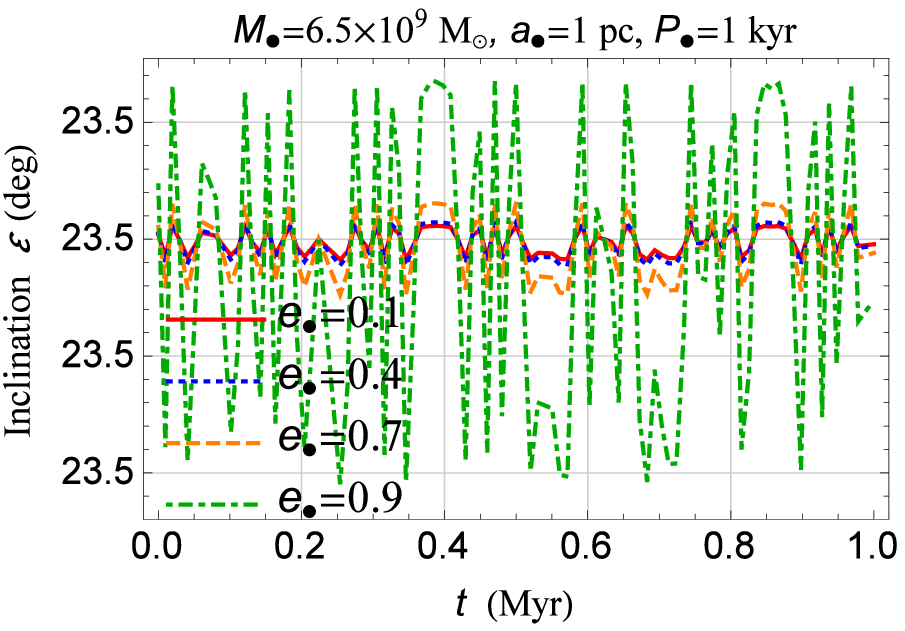}\\
\end{tabular}
}
}
\caption{
Upper panel: plot of the obliquity $\varepsilon$, in degrees, as a function of time $t$, in millions of years, over a time span $\Delta t=1\,\mathrm{Myr}$ for different values of the eccentricity $e_\bullet$ of the orbital motion of the star--planet system about an SMBH with $M_\bullet = 6.5\times 10^9\,M_\odot$ occurring in $P_\bullet = 1\,\mathrm{kyr}\ \ton{a_\bullet = 1\,\mathrm{pc}}$. In order to obtain $\Delta \varepsilon\ton{t}$, we straightforwardly integrated \rfr{obli} by keeping the planetary spin and galactocentric orbital parameters entering it fixed  to $I_\bullet=45^\circ,\,\Gamma=0$. They maximize the right-hand side of \rfr{obli}, which turns out to be positive. Lower panel: same as in the upper panel, apart from the fact that $\Delta \varepsilon\ton{t}$ was integrated by assuming the same precession of the equinoxes of the Earth for the spin's azimuth angle $\beta$. For the terrestrial planet we adopted the terrestrial values $J_2=  1.08\times 10^{-3},\,R=6378\,\mathrm{km},\,S = 5.86\times 10^{33}\,\mathrm{Js},\,\varepsilon=23^\circ.5$.
  }\label{figura3}
\end{figure}

If we assume for the planet's obliquity the Earth's value of $\varepsilon\simeq 23^\circ.5$, it turns out that the secular change induced by the SMBH is negligible, as shown by Figure\,\ref{figura3}.
Indeed, even by maximizing the trigonometric factor of \rfr{obli} for $I_\bullet =45^\circ,\Gamma=0$ and assuming no precession of the equinoxes, the resulting change experienced by $\varepsilon$ amounts just to $\Delta\varepsilon\lesssim 0^\circ. 02$ over $1\,\mathrm{Myr}$ even for large galactocentric eccentricities. Allowing the spin's azimuthal angle $\beta$ to precess in $26\,\mathrm{kyr}$ does not essentially alter the situation.
In the case of the real Earth, Equations\,(31)-(32) of \citet{2004A&A...428..261L} show that $\varepsilon$ exhibits an average rate as little as $\simeq 0^\circ.002\, \mathrm{Myr}^{-1}$, with jumps of the order of $\simeq 0^\circ .4$, over $500\,\mathrm{Myr}$.
\section{Summary and Conclusions}\lb{fine}
We first analytically worked out, in a perturbative way, the long-term orbital perturbations experienced by a fictitious Earth-like planet orbiting its Sun-type parent star because of the gravitational pull of an  SMBH no less than hundreds to thousands of its Schwarzschild radii away, depending on the eccentricity $e_\bullet$ of the two-body system's galactocentric motion. We doubly averaged them over both the yearly planet's orbital period $\Pb$ and the time $P_\bullet$ required to describe a galactocentric orbital revolution. The resulting expressions are valid in an arbitrary coordinate system, and are not restricted to any specific orbital geometries of both the planet and the SMBH. For the sake of definiteness, we assumed the mass of $\mathrm{M87}^\ast$ for the SMBH, whose apparent motion around the restricted two-body system, characterized by any value of the eccentricity $e_\bullet$, has a period $P_\bullet = 1\,\mathrm{kyr}$.
In order to tentatively look  at the ability of the planet to sustain complex life over time spans millions of years long, we focused on its perihelion distance $q$ and orbital inclination $I$, which are recognized as the fundamental quantities characterizing the insolation of the Earth. We preliminarily investigated their temporal evolutions over a $\simeq \mathrm{Myr}$ timescale by considering the planet both isolated in such a way that its orbital elements stay fixed, and impacted by possible $N$-body gravitational interactions with any other planets which we modeled as for the Earth. It turned out that, by assuming the specific orbital configurations which maximize their amplitudes, both $q$ and $I$ may undergo notable variations over a few million years which would seriously affect the habitability of the hypothesized world, especially for large values of $e_\bullet$. Suffice to say that, for $e_\bullet\simeq 0.7-0.9$, the planet, starting from the current heliocentric distance of the Earth, may even impact its star in $\simeq 2-3\,\mathrm{Myr}$ also when possible $N$-body harmonic modulations of the otherwise purely linear trends of $q$ are also taken into account. For the same values of $e_\bullet$, $I$ may be shifted by $\simeq 5^\circ-30^\circ$ or so.

Then, we also looked at the long-term changes of the planet's spin axis $\bds{\hat{S}}$ due to the SMBH, with particular emphasis on the obliquity $\varepsilon$ to the ecliptic-like plane, which is another key parameter for the habitability of an Earth-like world. We obtained perturbatively general analytical expressions for the rates of change of the three components of $\bds{\hat{S}}$ averaged over $P_\bullet$ by keeping them constant in the integration. The rate of change of ${\hat{S}}_z$  allowed us to infer the long-term precession of the obliquity. It turned out that, contrary to $q$ and $I$, $\varepsilon$ is not notably affected by the SMBH. Indeed, by using the same physical parameters of the Earth and the spin-orbital configuration which maximizes $\dot\varepsilon$, we obtained a shift as little as $\simeq 0^\circ .02$ over $1\,\mathrm{Myr}$. Also, the modulations induced by a possible Earth-like, slow precession of the equinoxes did not essentially alter such a conclusion.

We stress that the scope of our paper is not a detailed paleoclimatological study of the real Earth or of any other existing alien ecosystem over time; thus, our necessarily simplified approach, which, for example, assumed a purely Keplerian trajectory of the star--planet system around the galactic SMBH,
is adequate for our illustrative goals.

Our results, although necessarily preliminary and just indicative of the overall picture, have, nonetheless, a broader validity in a specific sense. Indeed, they can be applied not only to any orbital configurations of both the planet and the SMBH, but also to quite different astronomical and astrophysical systems like, e.g.,  extrasolar planets and, say, triple systems made of compact stellar corpses like neutron stars and white dwarfs usually described in a coordinate system whose fundamental reference plane coincides with the plane of the sky. Moreover, they can be also applied to the hunt for Telisto/Planet Nine in our solar system by looking at its effects, not only on the orbits of the other known planets, but also on their spin axes  and of the Sun itself.

I am grateful to an anonymous referee for her/his helpful remarks.

\bibliography{MR_biblio,MS_binary_pulsar_bib,Gclockbib,semimabib,PXbib}{}

\begin{thebibliography}{47}
\expandafter\ifx\csname natexlab\endcsname\relax\def\natexlab#1{#1}\fi

\bibitem[{{Abell}, {Morrison} \& {Wolff}(1993){Abell}, {Morrison}, \&
  {Wolff}}]{bookglobular}
{Abell} G.~O., {Morrison} D., {Wolff} S.~C., 1993, {Exploration of the
  Universe}. 6th ed.; Philadelphia, PA: Saunders College Publishing

\bibitem[{{Akiyama} {et~al}\mbox{.}(2019{\natexlab{a}}){Akiyama}, {Alberdi},
  {Alef}, \& {et al.}}]{2019ApJ...875L...1E}
{Akiyama} K., {Alberdi} A., {Alef} W., {et al.}, 2019{\natexlab{a}}, ApJL, 875,
  L1

\bibitem[{{Akiyama} {et~al}\mbox{.}(2019{\natexlab{b}}){Akiyama}, {Alberdi},
  {Alef}, \& {et al.}}]{2019ApJ...875L...6E}
{Akiyama} K., {Alberdi} A., {Alef} W., {et al.}, 2019{\natexlab{b}}, \apjl,
  875, L6

\bibitem[{{Bailey}, {Batygin} \& {Brown}(2016){Bailey}, {Batygin}, \&
  {Brown}}]{2016AJ....152..126B}
{Bailey} E., {Batygin} K., {Brown} M.~E., 2016, AJ, 152, 126

\bibitem[{{Balbi} \& {Tombesi}(2017)}]{2017NatSR...716626B}
{Balbi} A., {Tombesi} F., 2017, NatSR, 7, 16626

\bibitem[{{Batygin} {et~al}\mbox{.}(2019){Batygin}, {Adams}, {Brown}, \&
  {Becker}}]{2019PhR...805....1B}
{Batygin} K., {Adams} F.~C., {Brown} M.~E., {Becker} J.~C., 2019, PhR, 805, 1

\bibitem[{{Bertotti}, {Farinella} \& {Vokrouhlick\'{y}}(2003){Bertotti},
  {Farinella}, \& {Vokrouhlick\'{y}}}]{Bertotti03}
{Bertotti} B., {Farinella} P., {Vokrouhlick\'{y}} D., 2003, Physics of the
  Solar System. Dordrecht: Kluwer

\bibitem[{{Binney} \& {Tremaine}(1987)}]{1987gady.book.....B}
{Binney} J., {Tremaine} S., 1987, {Galactic Dynamics}. Princeton, NJ: Princeton
  Univ. Press

\bibitem[{Contopoulos \& Efthymiopoulos(2011)}]{Contopoulos:2011}
Contopoulos G., Efthymiopoulos C., 2011, Scholarpedia, 6, 10670, revision
  \#91294

\bibitem[{{Davari}, {Capuzzo-Dolcetta} \& {Spurzem}(2019){Davari},
  {Capuzzo-Dolcetta}, \& {Spurzem}}]{2019arXiv190806298D}
{Davari} N., {Capuzzo-Dolcetta} R., {Spurzem} R., 2019, arXiv e-prints,
  arXiv:1908.06298

\bibitem[{{Dokuchaev}(2011)}]{2011CQGra..28w5015D}
{Dokuchaev} V.~I., 2011, CQGra, 28, 235015

\bibitem[{{Dokuchaev}(2012)}]{2012GrCo...18...65D}
{Dokuchaev} V.~I., 2012, GrCo, 18, 65

\bibitem[{{Dressing} {et~al}\mbox{.}(2010){Dressing}, {Spiegel}, {Scharf},
  {Menou}, \& {Raymond}}]{2010ApJ...721.1295D}
{Dressing} C.~D., {Spiegel} D.~S., {Scharf} C.~A., {Menou} K., {Raymond} S.~N.,
  2010, ApJ, 721, 1295

\bibitem[{{Forbes} \& {Loeb}(2018)}]{2018MNRAS.479..171F}
{Forbes} J.~C., {Loeb} A., 2018, MNRAS, 479, 171

\bibitem[{{Gonzalez}, {Brownlee} \& {Ward}(2001){Gonzalez}, {Brownlee}, \&
  {Ward}}]{2001Icar..152..185G}
{Gonzalez} G., {Brownlee} D., {Ward} P., 2001, Icar, 152, 185

\bibitem[{{Habibi} {et~al}\mbox{.}(2019){Habibi}, {Gillessen}, {Pfuhl},
  {Eisenhauer}, {Plewa}, {von Fellenberg}, {Widmann}, {Ott}, {Gao}, {Waisberg},
  {Baub{\"o}ck}, {Jimenez-Rosales}, {Dexter}, {de Zeeuw}, \&
  {Genzel}}]{2019ApJ...872L..15H}
{Habibi} M. {et~al.}, 2019, ApJL, 872, L15

\bibitem[{{Hogg}, {Quinlan} \& {Tremaine}(1991){Hogg}, {Quinlan}, \&
  {Tremaine}}]{1991AJ....101.2274H}
{Hogg} D.~W., {Quinlan} G.~D., {Tremaine} S., 1991, AJ, 101, 2274

\bibitem[{{Kane} \& {Deveny}(2018)}]{2018ApJ...864..115K}
{Kane} S.~R., {Deveny} S.~J., 2018, ApJ, 864, 115

\bibitem[{Kauffmann(2016)}]{kauffmann_2016}
Kauffmann J., 2016, in IAU Symp. 322, The Multi-Messenger Astrophysics of the
  Galactic Centre, {Crocker} R., {Longmore} S., {Bicknell} G., eds., Cambridge:
  Cambridge Univ. Press, pp. 75--84

\bibitem[{{Kilic}, {Raible} \& {Stocker}(2017){Kilic}, {Raible}, \&
  {Stocker}}]{2017ApJ...844..147K}
{Kilic} C., {Raible} C.~C., {Stocker} T.~F., 2017, ApJ, 844, 147

\bibitem[{{Kopeikin}, {Efroimsky} \& {Kaplan}(2011){Kopeikin}, {Efroimsky}, \&
  {Kaplan}}]{2011rcms.book.....K}
{Kopeikin} S., {Efroimsky} M., {Kaplan} G., 2011, {Relativistic Celestial
  Mechanics of the Solar System}. Weinheim: Wiley-VCH

\bibitem[{{Kopp} \& {Lean}(2011)}]{2011GeoRL..38.1706K}
{Kopp} G., {Lean} J.~L., 2011, GeoRL, 38, L01706

\bibitem[{{Kruijssen} {et~al}\mbox{.}(2019){Kruijssen}, {Dale}, {Longmore}, \&
  {et al.}}]{Kruij2019}
{Kruijssen} J.~M.~D., {Dale} J.~E., {Longmore} S.~N., {et al.}, 2019, MNRAS,
  484, 5734

\bibitem[{{Laskar}, {Joutel} \& {Boudin}(1993){Laskar}, {Joutel}, \&
  {Boudin}}]{1993A&A...270..522L}
{Laskar} J., {Joutel} F., {Boudin} F., 1993, A\&A, 270, 522

\bibitem[{{Laskar}, {Joutel} \& {Robutel}(1993){Laskar}, {Joutel}, \&
  {Robutel}}]{1993Natur.361..615L}
{Laskar} J., {Joutel} F., {Robutel} P., 1993, Natur, 361, 615

\bibitem[{{Laskar} {et~al}\mbox{.}(2004){Laskar}, {Robutel}, {Joutel}, \& {et
  al.}}]{2004A&A...428..261L}
{Laskar} J., {Robutel} P., {Joutel} F., {et al.}, 2004, A\&A, 428, 261

\bibitem[{{Lingam}, {Ginsburg} \& {Bialy}(2019){Lingam}, {Ginsburg}, \&
  {Bialy}}]{Lingam_2019}
{Lingam} M., {Ginsburg} I., {Bialy} S., 2019, ApJ, 877, 62

\bibitem[{{Linsenmeier}, {Pascale} \& {Lucarini}(2015){Linsenmeier}, {Pascale},
  \& {Lucarini}}]{2015P&SS..105...43L}
{Linsenmeier} M., {Pascale} S., {Lucarini} V., 2015, P$\&$SS, 105, 43

\bibitem[{{Mehrgan} {et~al}\mbox{.}(2019){Mehrgan}, {Thomas}, {Saglia}, \& {et
  al.}}]{2019arXiv190710608M}
{Mehrgan} K., {Thomas} J., {Saglia} R., {et al.}, 2019, ApJ, 887, 195

\bibitem[{{Morris}(1993)}]{1993ApJ...408..496M}
{Morris} M., 1993, ApJ, 408, 496

\bibitem[{{Murphy}, {Gebhardt} \& {Cradit}(2014){Murphy}, {Gebhardt}, \&
  {Cradit}}]{2014ApJ...785..143M}
{Murphy} J.~D., {Gebhardt} K., {Cradit} M., 2014, ApJ, 785, 143

\bibitem[{{Murray} \& {Dermott}(2000)}]{2000ssd..book.....M}
{Murray} C.~D., {Dermott} S.~F., 2000, {Solar System Dynamics}. Cambridge:
  Cambridge Univ. Press

\bibitem[{{Opatrn{\'y}}, {Richterek} \& {Bakala}(2017){Opatrn{\'y}},
  {Richterek}, \& {Bakala}}]{2017AmJPh..85...14O}
{Opatrn{\'y}} T., {Richterek} L., {Bakala} P., 2017, AmJPh, 85, 14

\bibitem[{{Peebles}(1972)}]{1972ApJ...178..371P}
{Peebles} P.~J.~E., 1972, ApJ, 178, 371

\bibitem[{{Poisson} \& {Will}(2014)}]{2014grav.book.....P}
{Poisson} E., {Will} C.~M., 2014, {Gravity}. Cambridge: Cambridge Univ. Press

\bibitem[{{Prantzos}(2008)}]{2008SSRv..135..313P}
{Prantzos} N., 2008, SSRv, 135, 313

\bibitem[{{Schnittman}(2019)}]{2019arXiv191000940S}
{Schnittman} J.~D., 2019, arXiv e-prints, arXiv:1910.00940

\bibitem[{{Sch{\"o}del} {et~al}\mbox{.}(2002){Sch{\"o}del}, {Ott}, {Genzel}, \&
  {et al.}}]{2002Natur.419..694S}
{Sch{\"o}del} R., {Ott} T., {Genzel} R., {et al.}, 2002, Natur, 419, 694

\bibitem[{{Shemmer} {et~al}\mbox{.}(2004){Shemmer}, {Netzer}, {Maiolino}, \&
  {et al.}}]{2004ApJ...614..547S}
{Shemmer} O., {Netzer} H., {Maiolino} R., {et al.}, 2004, ApJ, 614, 547

\bibitem[{{Sloan}, {Alves Batista} \& {Loeb}(2017){Sloan}, {Alves Batista}, \&
  {Loeb}}]{2017NatSR...7.5419S}
{Sloan} D., {Alves Batista} R., {Loeb} A., 2017, NatSR, 7, 5419

\bibitem[{{Spiegel} {et~al}\mbox{.}(2010){Spiegel}, {Raymond}, {Dressing},
  {Scharf}, \& {Mitchell}}]{2010ApJ...721.1308S}
{Spiegel} D.~S., {Raymond} S.~N., {Dressing} C.~D., {Scharf} C.~A., {Mitchell}
  J.~L., 2010, ApJ, 721, 1308

\bibitem[{{Trani} {et~al}\mbox{.}(2016){Trani}, {Mapelli}, {Spera}, \&
  {Bressan}}]{2016ApJ...831...61T}
{Trani} A.~A., {Mapelli} M., {Spera} M., {Bressan} A., 2016, ApJ, 831, 61

\bibitem[{{Varadi}, {Runnegar} \& {Ghil}(2003){Varadi}, {Runnegar}, \&
  {Ghil}}]{2003ApJ...592..620V}
{Varadi} F., {Runnegar} B., {Ghil} M., 2003, ApJ, 592, 620

\bibitem[{{Vieira} {et~al}\mbox{.}(2012){Vieira}, {Norton}, {Dudok de Wit}, \&
  {et al.}}]{Vieira2012}
{Vieira} L.~E.~A., {Norton} A., {Dudok de Wit} T., {et al.}, 2012, GeoRL, 39,
  L16104

\bibitem[{{Wada}, {Tsukamoto} \& {Kokubo}(2019){Wada}, {Tsukamoto}, \&
  {Kokubo}}]{2019arXiv190906748W}
{Wada} K., {Tsukamoto} Y., {Kokubo} E., 2019, ApJ, 886, 107

\bibitem[{{Williams} \& {Kasting}(1997)}]{1997Icar..129..254W}
{Williams} D.~M., {Kasting} J.~F., 1997, Icar, 129, 254

\bibitem[{{Williams} \& {Pollard}(2002)}]{2002IJAsB...1...61W}
{Williams} D.~M., {Pollard} D., 2002, IJAsB, 1, 61

\end{thebibliography}


\end{document}